\documentclass[aps,twocolumn,prl,amsmath,amssymb,superscriptaddress,
showpacs,10pt]{revtex4-1} 
\usepackage{amsmath} \usepackage{amsfonts} \usepackage{amssymb}
\usepackage{amsthm} \usepackage{mathtools} \usepackage{mathrsfs}
\usepackage{color} \usepackage{xspace}
\usepackage[colorlinks, urlcolor=black, citecolor=blue, link
color=black]{hyperref} 

\def \Re{\text{Re}}

\def \Br{\text{Br}}
\def \Amp{\text{Amp}}

\def\Dbar {\kern 0.2em\overline{\kern -0.2em D}{}\xspace}

\def\Dzb   {\ensuremath{\Dbar^0}\xspace}
\def\DzDzb {\ensuremath{D^0 {\kern -0.16em \Dzb}}\xspace}

\def\Dz    {\ensuremath{D^0}\xspace}
\def\Dzzb {\kern 0.2em\overset{(\overline{\kern -0.2em D})}{}\xspace}

\def\Ds     {\ensuremath{D^*}\xspace}
\def\Dbars  {\ensuremath{\Dbar^*}\xspace}
\def\Dbs    {\ensuremath{\Dbars}\xspace}
\def\Dzbs   {\ensuremath{\Dbar^{*0}}\xspace}
\def\DzsDzbs {\ensuremath{D^{*0} {\kern -0.16em \Dzbs}}\xspace}

\def\Dzs    {\ensuremath{D^{*0}}\xspace}

\def\B     {\ensuremath{B}\xspace}

\def\Bbar  {\kern 0.18em\overline{\kern -0.18em B}{}\xspace}

\def\Bzb   {\ensuremath{\Bbar^0}\xspace}
\def\BzBzb {\ensuremath{B^0 {\kern -0.16em -\Bzb}}\xspace}

\def \CP 	{\ensuremath{C\!P}\xspace}

\def\K     {\ensuremath{K}\xspace}
\def\KS    {\ensuremath{K_{\!\scriptscriptstyle S}}\xspace} 
 
\def\Kbar  {\kern 0.2em\overline{\kern -0.2em K}{}\xspace}

\def\Kp    {\ensuremath{K^+}\xspace}
\def\Km    {\ensuremath{K^-}\xspace}

\def\Bbar  {\kern 0.18em\overline{\kern -0.18em B}{}\xspace}

\def\Bzb   {\ensuremath{\Bbar^0}\xspace}
\def\BzBzb {\ensuremath{B^0 {\kern -0.16em -\Bzb}}\xspace}

\def\nn    {\nonumber}
\def\sss{\scriptscriptstyle}

\def\pip    {\ensuremath{\pi^+}\xspace}
\def\pim    {\ensuremath{\pi^-}\xspace}

\def\piz    {\ensuremath{\pi^0}\xspace}

\newcommand{\braket}[2]{\left\langle #1 \right| \left. #2
\right\rangle}

\newcommand{\ket}[1]{\left| #1 \right\rangle}

\newcommand{\Ket}[1]{\big| #1\big\rangle}

\newcommand{\modulus}[1]{\left| #1 \right|}

\def\barp{{\raise.35ex\hbox{${\sss (}$}}---{\raise.35ex\hbox{${\sss )}$}}}
\def\bdbarp{\hbox{$B_d$\kern-1.4em\raise1.4ex\hbox{\barp}}}
\def\Dbarp{\hbox{$D$\kern-.85em\raise1.2ex\hbox{{{\raise.35ex
          \hbox{{\tiny(}}}--{\raise.35ex\hbox{${\sss )}$}}}}}}  
\def\deltabarp{\hbox{$\delta$\kern-.65em\raise1.2ex\hbox{{{\raise.35ex
          \hbox{{\tiny (}}}--{\raise.35ex\hbox{${\sss )}$}}}}}}

\def\sss{\scriptscriptstyle}

\def\Bbar{{\overline{B}}}

\def\barp{{\raise.35ex\hbox{${\sss (}$}}---{\raise.35ex\hbox{${\sss )}$}}}
\def\bdbarp{\hbox{$B_d$\kern-1.4em\raise1.4ex\hbox{\barp}}}
\def\bsbarp{\hbox{$B_s$\kern-1.4em\raise1.4ex\hbox{\barp}}}

\def\roughly#1{\mathrel{\raise.3ex\hbox{$#1$\kern-.75em
      \lower1ex\hbox{$\sim$}}}} 
\def\barpk{{\raise.35ex\hbox{${\sss (}$}}--{\raise.35ex\hbox{${\sss )}$}}}
\def\bbarp{\hbox{$B$\kern-0.9em\raise1.4ex\hbox{\barpk}}}
\def\Kstarzero{\widehat{K^*}{\raise0.75ex\hbox{\scriptsize 0}}}
\def\Kstarplus{\widehat{K^*}^{\lower0.75ex\hbox{\tiny +}}}

\def\bBar#1{\hbox{$#1$\kern -1.85em\raise1.6ex\hbox{{\raise.35ex\hbox{$~{\sss
(}$}}$-${\raise.35ex\hbox{${\sss )}$}}}\kern 0.3em}}

\allowdisplaybreaks

\begin{document}

\title{A new technique to observe direct
  \texorpdfstring{\boldmath{$\CP$}}{CP} Violation in
  \texorpdfstring{\boldmath{$D$}}{D} mesons\\using Bose Symmetry and
  Dalitz Plot}

\author{Dibyakrupa Sahoo} \affiliation{The Institute of Mathematical
  Sciences, Taramani, Chennai 600113, India}
\author{Rahul Sinha} \affiliation{The Institute of Mathematical
  Sciences, Taramani, Chennai 600113, India}
\author{N.\ G.\ Deshpande} \affiliation{Institute of Theoretical
  Science, University of Oregon, Eugene, OR 94703, USA}
\author{Sandip Pakvasa} \affiliation{Department of Physics and
  Astronomy, University of Hawaii, Honolulu, HI 96822, USA}

\date{\today}

\begin{abstract}
  We present a new and sensitive method to observe direct $\CP$
  violation in $D$ mesons using Bose symmetry and Dalitz plot. We
  apply the method to processes such as $B\to \Dz\Dzb P$, where $P$ is
  either a $K$ or a $\pi$. By choosing to reconstruct $D$ mesons only
  through their decays into $\CP$ eigenstates, we show that any
  asymmetry in the Dalitz plot can arise only through direct $\CP$
  violation. We further show how $\CP$ violation parameters can be
  determined. Since the approach involves only Bose symmetry, the
  method is applicable to any multi-body process that involves
  $\Dz\Dzb$ in the final state. We briefly discuss how $B\to \Ds\Dbs
  P$ can also be used in a similar way.
\end{abstract}


\maketitle

It is well known that the explanation of observed predominance of
matter over antimatter in our universe requires as one of the
conditions that the laws of nature are not invariant under $\CP$
\cite{ref:Sakharov}. Indeed $\CP$ violation has been well established
in the weak decays of $K$ \cite{ref:Christenson1964} and $B$ mesons
\citep{ref:Aubert2001, ref:Abe2001, ref:Aubert2004, ref:Abe2004,
  ref:Poluektov2010, ref:Sanchez2010, ref:Aaij2012a}. In the standard
model of particle physics $\CP$ violation arises via the
Kobayashi-Maskawa mechanism at a level consistent with that observed
in the $K$ mesons and $B$ mesons. However, it is well known that $\CP$
violation in the standard model cannot account for the observed baryon
asymmetry making it imperative to search for new sources of $\CP$
violation beyond the standard model. In the standard model, $\CP$
violation in the $D$ meson system is expected to be rather small in
both mass mixing and in direct decays. An observation of sizable $\CP$
violation in $D$ mesons would hence open a window of opportunity to
probe for new sources of $CP$ violation. This in turn may lead to a
more complete theory of $CP$ violation that furthers our understanding
of the observed baryon asymmetry. It is, however, challenging to
observe an unambiguous signal of $\CP$ violation in $D$ mesons.  In
this letter we present a new technique to observe direct $\CP$
violation using Bose symmetry and Dalitz plots. The method is
completely general but we apply it here to study $\CP$ violation in
$D$ mesons.

Our approach relies on using $D$ mesons produced in decays such as
$B\to\Dz\Dzb K$ or $B\to \Dz\Dzb\pi$ or any similar mode such as $B\to
\Ds\Dbs P$ that result in $\Dz\Dzb$ in the final decay chain. The
$\Dz\Dzb$ pair is best viewed in terms of the mass eigenstates $D_1$
and $D_2$ so that the time evolution is simple. The final $\Dz\Dzb$
state can be observed as a combination of $D_1 D_1$, $D_2 D_2$ and
$D_1D_2$ states.  In the absence of $\CP$ violation in the mixing
matrix of $\Dz$-$\Dzb$ system, $D_1$ and $D_2$ would be even and odd
eigenstates of $\CP$. In the three body Dalitz plot because of Bose
symmetry, we necessarily have $D_1D_1$ and $D_2D_2$ in symmetric
states under exchange of momenta. Further, as we will see, a $D_1D_2$
state is necessarily anti-symmetric under the exchange of momenta. Now
if there is no $\CP$ violation in both mass-mixing (indirect $CP$
violation) and directly in the decay amplitude, and if one observes
both $D$ mesons reconstructed in $\CP$ even eigenstates (such as
$\Kp\Km$ or $\pip\pim$), this can come only from the $D_1D_1$, and
hence would be distributed symmetrically under the exchange of the two
identical D mesons in the Dalitz plot. The same is true if we
reconstruct both the $D$ mesons in $\CP$ odd eigenstates such as $\KS
\pi^0$, $\KS\omega$ or $\KS\phi$, which arise from decays of $D_2$.

The advantage of studying Dalitz plot distributions with due care to
the consequences of Bose symmetry has also been eluded to
earlier~\cite{Sinha:2011ky}.  The crucial observation of this letter
is that any asymmetry in the Dalitz plot distribution of $B\to \Dz\Dzb
P$ events under exchange of the $D$ mesons, when the $D$'s are
reconstructed in same $\CP$ eigenstates, must necessarily arise from
the antisymmetric state $D_1D_2$ and imply $\CP$ violation, since it
would mean that both $D_1$ and $D_2$ decay into the same $\CP$ final
state. We prove that an observation of such an asymmetry would
constitute an unambiguous signal of {\em direct} $\CP$ violation which
is independent of $\CP$ violation in mass mixing.  Further, the
asymmetry in the Dalitz plot would arise from the interference of
$\CP$ violating and $\CP$ conserving terms. Hence, our procedure could
be more sensitive to $\CP$-violation compared to the observation of
$\CP$-violation using measurement of differences in decay rates. The
novel observation made is that the distribution of points on the
Dalitz plot depends on the $\CP$-property of the decay mode chosen to
reconstruct the $D$ mesons. This follows as a consequence of the Bose
symmetry of final states in the Dalitz plot distribution.  The
technique presented here promises to be a powerful new tool for
discovery of new physics.

The mass eigenstates of the two neutral $D$ mesons are defined as
\begin{equation}
\ket{D_{1,2}} = p \ket{\Dz} \pm q \ket{\Dzb},
\end{equation}
where, $\modulus{p}^2 + \modulus{q}^2 = 1$, and the two $\CP$
eigenstates are
\begin{equation}
\ket{D_{\pm}} = \frac{1}{\sqrt{2}} 
\Big( \ket{\Dz} \pm \ket{\Dzb} \Big).
\end{equation}
The $\Dz$ and $\Dzb$ mesons can then be expressed in terms of the mass
eigenstates and the $\CP$ eigenstates as follows:
\begin{equation} \label{eq:Dz-Dzbar}
\Ket{\bBar{\Dz}}= \frac{1}{2p} \Big( \ket{D_1} \pm \ket{D_2} \Big)
= \frac{1}{\sqrt{2}} \Big( \ket{D_{+}} \pm \ket{D_{-}} \Big).
\end{equation}
Finally the mass eigenstates can be written in terms of the $\CP$
eigenstates at any given time as follows:
\begin{equation} \label{eq:D12}
\ket{ D_{1,2} } = \frac{1}{\sqrt{2}} 
\Big( \left( p \pm q \right) \ket{D_{+}} +
\left( p \mp q \right) \ket{D_{-}} \Big).
\end{equation}
The states $\ket{ D_{1,2} }$ have only an exponential time dependence
corresponding to their mass and decay width and do not depend on time
in any other way.  This time dependence is given by
\begin{align}\label{eq:time-dependence}
\ket{ D_{1,2}(t)} &=e^{-i\mu_{1,2} t}\ket{D_{1,2}} \equiv
e^{-i(\mu\pm\Delta\mu)t}\ket{D_{1,2}},  
\end{align}
where $\mu= M- i \, \left( \Gamma / 2 \right)$ and $\Delta\mu = (x - i
\, y) \, \left( \Gamma / 2 \right)$ with $M$ and $\Gamma$ being the
average mass and decay width of $D_1$ and $D_2$ and $x \Gamma$, $2 \,
y \, \Gamma$ being the differences in masses and widths of $D_1$ and
$D_2$ respectively. Experimentally~\cite{ref:PDG} $x =
(0.48^{+0.17}_{-0.19}) \times 10^{-2}$ and $y=(0.715 \pm 0.095)\times
10^{-2}$.  We choose to work in terms of mass eigenstates as the time
dependence is a simple exponential that can be easily integrated, and
these states also obey Bose statistics.

We consider the decay $B\to\Dz(p_1)\Dzb(p_2)P(p_3))$ in the
Gottfried-Jackson frame with the $B$ moving along the $\hat{z}$ axis
so that the $D^0$ and $\overline{D}^0$ go back to back with $D^0$ at
an angle $\theta$ with $P(p_3)$.  In this frame
$\vec{p_1}+\vec{p_2}=0$. We define $s \equiv (p_1 +p_2)^2$, $t \equiv
(p_2 +p_3)^2$ and $u \equiv (p_1 +p_3)^2$. Here $t$ and $u$ can be
written as:
\begin{equation} \label{eq:tu}
t \equiv a+b \,\cos\theta~, \qquad u \equiv a-b \,\cos\theta~, 
\end{equation}
where $a = \left( M_B^2 + M^2_P+ 2 M^2_{D}-s \right)/2$ and $b =
\sqrt{s -4 M^2_{D}} \; \lambda^{1/2}(M_B^2,M^2_P,s) / \left( 2\sqrt{s}
\right)$, with $\lambda(x,y,z)=x^2 + y^2 + z^2 - 2 (xy + yz + zx)$.
The $\Dz$ and $\Dzb$ mesons produced in the decay of the $B$ meson at
time $t=0$ oscillate and eventually decay by weak interactions to the
final states $f_1$ and $f_2$ at times $t_1$ and $t_2$ respectively.
We shall first consider the Dalitz plot for the decay $B\to \Dz\Dzb
P$, where both the $D$ are reconstructed in $\CP$-even states. Since
the two $D$ mesons are reconstructed in $\CP$ eigenstates we cannot
ascertain the flavor of the $D$ meson. Hence, we choose the D meson
reconstructed in the final state $f_1^{+}$ to have momentum $p_1$ and
the other $D$ meson with momentum $p_2$ is presumed to decay to
$f_2^{+}$. The superscript $+$ is used to indicate that the state is a
$\CP$-even eigenstate. Here $f_1^{+}$ and $f_2^{+}$ could be for
example $\pi^+\pi^-$, $K^+K^-$ or any other $\CP$-even final state. To
obtain an asymmetry it is necessary that $f_{1}^{+}\neq f_2^{+}$. In
the case $f_{1}^{+}= f_2^{+}$ there is a two fold ambiguity in
assigning a point on the Dalitz plot. These pair of points are related
by $t\leftrightarrow u$, and we can plot the points uniquely in the
half plane $t\geqslant u$.  We will show later that this plot contains
important information.

Our interest lies in the decays of the $\Dz$ and $\Dzb$ mesons
decaying as mass eigenstates. Hence we express the state
$\ket{\Dz(p_1) \Dzb(p_2) P} $ in terms of mass eigenstates of the
$\Dz$ and $\Dzb$ mesons (using Eq.~\eqref{eq:Dz-Dzbar}) as follows:
\begin{align}
\label{eq:Dz-Dzb-state1}
& \ket{\Dz(p_1) \Dzb(p_2) P} = \nn\\ 
& \frac{1}{4pq} \Big( \Big\{\ket{D_1(p_1) D_1(p_2)P} - \ket{D_2(p_1)
  D_2(p_2)P} \Big\} \nn\\ 
& \phantom{***} - \Big\{ \ket{D_1(p_1) D_2(p_2)P} -
\ket{D_2(p_1) D_1(p_2)P} \Big\} \Big).
\end{align}
where the momenta of the two $D$ mesons have been explicitly written
to emphasize the exchange symmetry between the $D_1$ and $D_2$
mesons. Since the two mesons are reconstructed in $\CP$-eigenstates
that do not tag the flavor of the $\Dz$ and $\Dzb$ meson, the symmetry
properties of the two $D$ mesons need to be taken care off. It is
clear from Eq.~\eqref{eq:Dz-Dzb-state1} that the term in the first
curly bracket is symmetric under the exchange of the momenta of the
two $D$ mesons, whereas the second term is antisymmetric under the
same exchange.  Exchanging $p_1$ and $p_2$ we obtain the expression
for the state $\ket{\Dz(p_2)\Dzb(p_1)P}$ which is identical to the
right-hand-side of Eq.~\eqref{eq:Dz-Dzb-state1} except a change of
sign in front of the second curly bracket.  The term symmetric under
exchange of $p_1$ and $ p_2$ exhibits Bose symmetry as it has
identical mesons. Since, the exchange $p_1 \leftrightarrow p_2$ is
equivalent to the exchange $t \leftrightarrow u$, the Bose symmetry is
therefore realized as a symmetry under $t \leftrightarrow u$ exchange
in the amplitude and consequently in the Dalitz plot.

We define the amplitude for the decay of $D_{\pm}$ to a $\CP$ even
final state $f_i^+$ as
\begin{align}
\label{A_ip}
\Amp(D_{+}\to f_i^{+})&=\braket{f_i^+}{D_{+}}=A_i\\
\label{A_im}
\Amp(D_{-}\to f_i^{+})&=\braket{f_i^+}{D_{-}}=\epsilon_i A_i,
\end{align}
where $\epsilon_i$ clearly indicate $\CP$ violation. The amplitude for
the decay $D_{1,2}\to f_i^{+}$ can be written using Eq.~\eqref{eq:D12}
as,
\begin{equation}
\Amp(D_{1,2}\to f_i^{+})=\frac{1}{\sqrt{2}}\Big((p\pm q)\,A_i+(p\mp q)
\epsilon_i \, A_i \Big).
\end{equation}

In order to understand the nature of $\CP$ violation possible in our
method we first examine two simple cases in the limit of vanishing
mass and width differences between $D_1$ and $D_2$:
\paragraph{\textbf{Case (a) No direct $\CP$ violation}}
In this limit we have $\epsilon_i=0$, resulting in a simpler
expression for the amplitude for the decay $D_{1,2}\to f_i^{+}$:
\begin{equation}
\Amp(D_{1,2}\to f_i^{+})=\frac{1}{\sqrt{2}}(p \pm q)\,A_i.
\end{equation}
It is obvious that the contribution from both the states
$\ket{D_1(p_1) D_2(p_2)P}$ and $\ket{D_2(p_1) D_1(p_2)P}$ decaying to
$f_1^{+}f_2^{+}P$ is each proportional to $(p^2-q^2)$. Hence, the odd
state $\big\{\ket{D_1(p_1) D_2(p_2)P} - \ket{D_2(p_1) D_1(p_2)P}
\big\}$ gives two contributions that exactly cancel. Thus no asymmetry
would be observed in the Dalitz plot distribution if direct $\CP$
violation were absent.
\paragraph{\textbf{Case (b) No $\CP$ violation in mixing}}
In this case we take the limit $p=q$ and hence
\begin{equation}
\ket{D_{1,2}}=\ket{D_\pm}
\end{equation}
Using Eq.~\eqref{A_ip} and \eqref{A_im}, it is easy to see that the
odd-state $\big\{\ket{D_1(p_1) D_2(p_2)P} - \ket{D_2(p_1) D_1(p_2)P}
\big\}$ has a contribution proportional to $(\epsilon_i-\epsilon_j)
A_i A_j$.  Hence, we conclude that only direct $\CP$ violation can
contribute to the asymmetry observed in the Dalitz plot.

We now discuss the completely general case where no assumption is made
about the nature of $\CP$ violation.  Since, the mass difference $(x
\, \Gamma)$ and width difference ($2 \, y \, \Gamma$) of $D_1$ and
$D_2$ are negligible compared to the decay width ($\Gamma$), we shall
ignore these differences by setting $x\to 0$ and $y \to
0$. Corrections due to retaining $x$ and $y$ are extremely
small. Ignoring the mass and width differences, the amplitude for the
$B$ meson decay to the exchange-symmetric (even) final state $\big\{
(f_1^{+})_{D_1} (f_2^{+})_{D_1}P \, - \, (f_1^{+})_{D_2}
(f_2^{+})_{D_2}P \big\}$, is given by
\begin{widetext}
\begin{equation} \label{eq:ampeven}
\Amp\Big(B\to \big\{(f_1^{+})_{D_1} (f_2^{+})_{D_1} P-(f_1^{+})_{D_2}
(f_2^{+})_{D_2}P\big\}\Big) = 2 \,A_1 A_2 \, e^{-i\mu (t_1+t_2)} \big(
A(t,u) + A(u,t) \big) \, p \, q \, (1-\epsilon_1\epsilon_2),
\end{equation}
where, $(f_i^{+})_{D_1}$ indicates that $D_1$ decays to the final
state $f_i^{+}$, $A(t,u)$ is a function of $t$ and $u$ that describes
the amplitude for the decay of the $B$ meson to $\Dz\Dzb P$. One can
similarly write the amplitude for the $B$ meson decay to the
exchange-antisymmetric (odd) final state $\big\{ (f_1^{+})_{D_1}
(f_2^{+})_{D_2}P \, - \, (f_1^{+})_{D_2} (f_2^{+})_{D_1}P \big\}$ as
\begin{equation} \label{eq:ampodd}
\Amp\Big(B\to \big\{(f_1^{+})_{D_1} (f_2^{+})_{D_2} P-(f_1^{+})_{D_2}
(f_2^{+})_{D_1}P\big\}\Big) = 2 \, A_1 A_2 \, e^{-i\mu (t_1+t_2)} 
\big( A(t,u) - A(u,t) \big) \, p \, q \, (\epsilon_2-\epsilon_1)~.
\end{equation}
The complete amplitude for the decay $B \to (f_1^{+})_{D}
(f_2^{+})_{D} P$ is given by the sum of the amplitudes as given in
Eqs.~\eqref{eq:ampeven} and \eqref{eq:ampodd}:
\begin{equation} \label{eq:leadingorder}
\Amp(B \to (f_1^{+})_{D} (f_2^{+})_{D} P)= A_1 \, A_2 \, e^{-i \mu
  (t_1 + t_2)} \Big(A_e (1-\epsilon_1 \epsilon_2) - A_o (\epsilon_1 -
\epsilon_2) \, \cos\theta \Big)~.
\end{equation}
where $A_e$ and $A_o$ are both even under $t \leftrightarrow u$ or
$\cos\theta \leftrightarrow -\cos\theta$ (see Eq.~\eqref{eq:tu}) and
are defined as:
\begin{equation}
A_e=\frac{A(t,u)+A(u,t)}{2}~,\qquad 
A_o=\frac{A(t,u)-A(u,t)}{2\cos\theta}~.
\end{equation}
The Dalitz plot distribution for $B\to (f_1^{+})_{D} (f_2^{+})_{D} P$
is proportional to the time integrated differential decay rate.  We
find this to be:
\begin{align}
\label{eq:Dalitz-dist}
\frac{d\Gamma\big(B\to (f_1^{+})_{D} (f_2^{+})_{D} P\big)}{dt \, du}
&= \frac{\Br_1^+ \, \Br_2^+}{256 \, \pi^3 \, M_B^3}
\modulus{1+\epsilon_1}^{-2} \modulus{1+\epsilon_2}^{-2} \, \Big(
\modulus{1-\epsilon_1\epsilon_2}^2
\modulus{A_e}^2+\modulus{\epsilon_1-\epsilon_2}^2\modulus{A_o}^2 
\cos^2\theta \nn\\
& \phantom{***********************}
-2\,\Re\big((\epsilon_1-\epsilon_2)\,(1-\epsilon_1^* \epsilon_2^*)
A_e^* A_o\big)\cos\theta \Big),
\end{align}
\end{widetext}
where $\Br_1^+$ and $\Br_2^+$ are the branching fractions for the
processes $\Dz \to f_1^+$ and $\Dz \to f_2^+$ respectively. Since $t-u
\propto \cos\theta$, it is evident from Eq.~\eqref{eq:Dalitz-dist}
that any asymmetry across the line $t=u$ is an unambiguous signature
of direct $\CP$ violation in $D$ mesons.  We emphasize that this
method of observing direct $\CP$ violation in $D$ mesons is in no way
related to the traditional method of comparing Dalitz plots of $B$
decay and its conjugate process, which in this case would indicate
direct $\CP$ violation in $B$ decays and would be insensitive to any
$\CP$ violation in $D$ decays. Another approach~\cite{Giri:2003ty}
which is completely unrelated to ours, has previously used the Dalitz
plot of $D$ decay to observe evidence of $\CP$ violation in a $B$
decay (e.g. $B^+ \to D \Kp, D \to K \pi \pi$).

In the above we have discussed in detail only the final states where
both the meson decay to $\CP$-even eigenstates. The same analysis can
also be carried out when both the $D$ mesons decay to $\CP$-odd states
$f_1^{-}$ and $f_2^{-}$, where $f_1^{-}$ and $f_2^{-}$ are one of the
following $\KS \piz$, $\KS \omega$ or $\KS \phi$. It is easy to verify
that the differential decay width $D^{--}_{1,2}$ describing the
distribution for the process $B\to (f_1^{-})_{D} (f_2^{-})_{D} P$ is
exactly the same as the expression for $D^{++}_{1,2}$ in
Eq.~\eqref{eq:Dalitz-dist}, with the replacements $A_i=\Amp(D_{-}\to
f_i^{-})=\braket{f_i^-}{D_{-}}$ and $\epsilon_i \to \varepsilon_i=
\Amp(D_+\to f_i^-)/\Amp(D_-\to f_i^-)$. Note, that $\varepsilon$
($\epsilon$) denotes $\CP$ violation for states that are $\CP$-odd
(even).

Another interesting possibility is to reconstruct the Dalitz plot in a
mixture of $\CP$-even and $\CP$-odd states i.e. $f_i^{-} f_j^{+}$. If
$\CP$ is conserved this would contribute only to the anti-symmetric
part of the amplitude, but nevertheless the Dalitz plot would be
symmetric across $t=u$, since the Dalitz plot is proportional to the
mod-squared of the amplitude. In the presence of direct
$\CP$-violation the amplitude develops a symmetric part, once again
leading to an asymmetry in the Dalitz plot across $t=u$. The
difference in this case is that the coefficients of the expressions in
Eq.~\eqref{eq:ampeven} and \eqref{eq:ampodd} are interchanged. This
results in the following expression for the differential decay width
\begin{widetext}
\begin{align}
\label{eq:Dalitz-distfmfp}
\frac{d\Gamma\big(B\to (f_1^{-})_{D} (f_2^{+})_{D} P)\big)}{dt \, du}
&= \frac{\Br_1^- \, \Br_2^+}{256 \, \pi^3 \, M_B^3}
\modulus{1+\varepsilon_1}^{-2} \modulus{1+\epsilon_2}^{-2} \, \Big(
\modulus{1-\varepsilon_1\epsilon_2}^2 \modulus{A_o}^2 +
\modulus{\varepsilon_1-\epsilon_2}^2\modulus{A_e}^2 \cos^2\theta \nn\\
& \phantom{***********************}
-2\,\Re\big((\varepsilon_1-\epsilon_2)\,(1 - \varepsilon_1^*
\epsilon_2^*) A_o^* A_e\big)\cos\theta \Big)~.
\end{align}
\end{widetext} 

The Dalitz plot distribution can be written in the form
$E\,+\,O\,\cos\theta$, where the $E$ and $O$ are defined as the even
and odd parts of the distribution. These terms are distinguished by
their behavior under $t \leftrightarrow u$.  The odd term is the
coefficient of $\cos\theta$ which changes sign under $u
\leftrightarrow t $, however both $E$ and $O$ are by themselves even
functions of $\cos\theta$.  It is easy to see from
Eqs.~\eqref{eq:Dalitz-dist} that when $f_1^+=f_2^+$ in the final
state, only $E$ term proportional to $\modulus{A_e}^2$
survives. Hence, $\modulus{A_e}^2$ can easily be extracted from such a
symmetric Dalitz plot distribution defined here as $E_{\rm sym}$. This
holds true even in the case $f_1^-=f_2^-$. It may be noted that
Eq.~\eqref{eq:Dalitz-distfmfp} enables us to measure $\modulus{A_o}^2$
using the even ($E$) term whereas the odd ($O$) term can then be used
to extract the $\CP$ violation parameter $\varepsilon_1-\epsilon_2$
modulo the relative phase.  The expressions for the $\CP$ violation
parameter $\varepsilon_1-\epsilon_2$ is thus given by
\begin{equation}\label{eq:diff-1} 
\modulus{\varepsilon_1 - \epsilon_2} \geqslant \frac{\Br_{\rm sym}
}{2\,\sqrt{\Br_1^- \Br_2^+}} \, \frac{O}{\sqrt{E_{\rm sym} \, E}}~,
\end{equation} 
where $\Br_{\rm sym}$ is the branching fraction for $\Dz$ to the final
state $f_1 = f_2$.  If we consider the odd ($O$) term from the Dalitz
distribution for $B \to (f_1^+)_D (f_2^+)_D P$, then the expression
for $\CP$ violation parameter $\epsilon_1 - \epsilon_2$ is given by
\begin{equation}\label{eq:diff-2} 
\modulus{\epsilon_1-\epsilon_2} \geqslant \frac{\Br_{\text{sym}}}{2
\, \Br_1^+} \, \sqrt{\frac{\Br_1^-}{\Br_2^+}} \; \frac{O}{\sqrt{E \,
E_{\text{sym}}}}~,
\end{equation}
where $\Br_1^-$ is the branching fraction for $\Dz \to f_1^-$, where
$f_1^-$ is the mode used to measure $A_o$ as described above.  In
deriving Eqs.~\eqref{eq:diff-1} and \eqref{eq:diff-2} we have
neglected higher powers of $\epsilon$'s and $\varepsilon$'s in
comparison with unity. The $\CP$ violating asymmetry
$\modulus{\epsilon_1 - \epsilon_2}$ (or $\modulus{\varepsilon_1 -
  \epsilon_2}$) is easily related to the difference in $\CP$
asymmetries usually referred to as $\Delta A_{\CP} = A_{\CP}(f_1^{+})
- A_{\CP}(f_2^+)$ (or $\Delta A_{\CP} = A_{\CP}(f_1^{-}) -
A_{\CP}(f_2^+)$)~\cite{ref:Aaij2012b}. Note that even though $E$,
$E_{sym}$ and $O$ are functions of $t$ and $u$, measurement of the
$\CP$ violation parameter can be done by integrating over regions of
$t$ and $u$. The easiest way to observe $\CP$ violation is to measure
the difference between the number of events with $t > u$ and those
with $t < u$ and compare this to the total number of events on the
Dalitz plot.

Having established that any asymmetry in the Dalitz plot distribution
of the three body decay $B\to \Dz\Dzb P$ indicates the presence of
$CP$ violation, we next examine the mode $B\to \Ds \Dbs P$, which has
a much higher branching fraction. The $\Ds$ and $\Dbs$ undergo further
decays resulting in a five body final state that includes
$\Dz\Dzb$. However, the presence of the additional two particles adds
only to the complication in reconstruction. If the $D$ mesons are
reconstructed in two different $\CP$-even eigenstates e.g. $f_1^+$,
$f_2^+$, the resulting $\Ds \Dbs P$ Dalitz plot should be symmetric
under $t\leftrightarrow u$ exchange in the absence of $\CP$
violation. This is easy to see since $f_1^+$ (or equivalently $f_2^+$)
could be produced by the decay of $D_1$ which in turn can arise from
the decay of either $\Ds$ or $\Dbs$. The inequalities of
Eqs.~\eqref{eq:diff-1} and \eqref{eq:diff-2} are also valid for the
$\Ds\Dbs P$ Dalitz plots. However, the $E$, $E_{\text{sym}}$ and $O$
terms should now be extracted from the even and odd parts of the
$\Ds\Dbs P$ Dalitz plots, while the branching fractions are still
those of the $D$ mesons.

We estimate the sensitivity of our approach by considering $\B^\pm \to
\Dzs \Dzbs \K^\pm\to (\Dz\piz)(\Dzb\piz)\K^\pm$ decay where the
$\Dz\Dzb$ pair is reconstructed via $(\KS\piz)(\Kp\Km)$. A sample of
$\sim 3\times 10^{11}$ $B$ decays would enable us to probe the $\CP$
violation parameter of the order of $10^{-2}$ at $5\sigma$.  If the
$\Dz\Dzb$ pair were reconstructed via $(\Kp\Km)(\pip\pim)$, the same
sensitivity would require $\sim 3\times 10^{12}$ $B$ decays.  These
estimates are based only on statistical errors and do not include
detection efficiencies. Combining various decay modes one can enhance
the reach significantly. The current measurements \cite{Aubert:2007if,
  ref:Aaij2012b, Collaboration:2012qw, Ko:2012jh, LHCb:2013dka,
  Ko:2013qva} suggest a central value of $\Delta
A_{\CP}=A_{\CP}(K\!K)-A_{\CP}(\pi\!\pi)$ to be about $-0.34\times
10^{-2}$ but are not yet significant.

In this letter we have presented a new method to observe direct $\CP$
violation in $D$ mesons using three body $B$ decays.  The method
relies only on Bose symmetry and requires that the Dalitz plot of $B$
decays be constructed with the $D$ mesons decaying to specific $CP$
eigenstates. The signature of \textit{direct} $\CP$ violation in the
$D$ mesons is the presence of an asymmetry in the number of events
under $t\leftrightarrow u$ on the Dalitz plot distribution.

\acknowledgments This work is supported in part by DOE under contract
numbers DE-FG02-96ER40969ER41155 and DE-FG02-04ER41291. We thank Tom
Browder for detailed discussions. R.S. thanks N.G.D. and Institute of
Theoretical Science, University of Oregon for hospitality. S.P.
acknowledges the support of Alexander von Humboldt Foundation Research
Award and Professor Heinrich Paes, Physics Department of the Technical
University of Dortmund for hospitality.

\end{document}